# Induced-quantum magnetism on a triangular lattice of non-Kramers ions in $PrMgAl_{11}O_{19}$

S. Kumar,[1,2], M. Klicpera,[1] A. Eliáš,[1] M. Kratochvílová,[1] A. Kancko,[1] C. Correa,[3] K. Załęski,[4] M. Śliwińska-Bartkowiak,[2] R. H. Colman,[1] and G. Bastien[1*]

[1] *Charles University, Faculty of Mathematics and Physics, Department of Condensed Matter Physics, Prague, Czech Republic*

[2] *Adam Mickiewicz University, Faculty of Physics and Astronomy, Department of Experimental Physics of Condensed Phase, Poznan, Poland*

[3] *Institute of Physics of the Czech Academy of Sciences, Na Slovance, Prague, Czech Republic*

[4] *Adam Mickiewicz University, NanoBioMedical Centre, Poznan, Poland*

**Corresponding author:* sonu.kumar@matfyz.cuni.cz

## Abstract

We report the magnetic properties of the quantum triangular lattice antiferromagnet (TLAF) $PrMgAl_{11}O_{19}$ through magnetization and specific heat measurements. Strong magnetic anisotropy indicates the realization of an Ising-like magnetism in $PrMgAl_{11}O_{19}$ single crystal while no long-range magnetic ordering is realized down to 0.4 K. The splitting of the low-lying quasi-doublet of the non-Kramers ion $Pr^{3+}$ into two singlets suggested by experimental data, is consistent with an effective pseudospin-1/2 scenario. Single crystal XRD shows the repartition of $Pr^{3+}$ ions between two slightly different positions in their oxygen cage and specific heat measurements consistently show a double Schottky anomaly, attributed to excitations between the two singlets of the low-lying quasi-doublet from the two distinct sites. The observed gapless excitations in zero field are attributed to induced quantum magnetism, predicted when the magnetic interactions are of an energy scale comparable with the splitting between the two singlets. Based on these results, we find that the magnetic ground state of $PrMgAl_{11}O_{19}$ can be modeled by a quantum Ising magnet with an intrinsic transverse field rather than a quantum spin liquid (QSL).

## Introduction

The exploration of quantum magnetism has emerged as a significant theme in condensed matter physics, driven by the quest to discover exotic states of matter that harbor non-local topological excitations or quantum entanglement. So far, research in quantum magnetism has largely focused on systems with spin or effective spin-1/2 moments, where the formation of quantum spin liquids (QSLs) is anticipated [1-4]. However, recent studies have revealed that frustrated magnets with integer spin S or total angular momentum *J* can also host a rich variety of quantum magnetic phases, including quantum Ising magnets [5-12], quantum nematic phases [13-15], quadrupolar orders [16-18], intertwined dipolar-multipolar orders [19-20], and disordered dipolar-multipolar states [21].

Frustrated magnets involving non-Kramers rare earth ions such as $Pr^{3+}$ and $Tm^{3+}$ with J = 4 and J = 6 are promising platforms for studying these exotic phases [6,10,11,22]. Unlike Kramers ions, where each crystal electric field (CEF) state is at least doubly degenerate, non-Kramers ions can have a split quasi-doublet ground state (two non-degenerate singlets), which individually carry no dipole moment but any mixture of these two levels carries a magnetic moment [5-7,10,22-25]. These two levels also carry multipolar moments [19,20], which can lead to the formation of an uncommon magnetic order combining dipolar with higher-order multipolar orders [19], such as those observed in the triangular magnet $TmMgGaO_4$ [20]. In $TmMgGaO_4$, the magnetic order forms upon a double phase transition including a topological Berezinskii-Kosterlitz-Thouless (BKT) phase transition [8,26-28].

Recently, the triangular magnets $PrZnAl_{11}O_{19}$ and $PrMgAl_{11}O_{19}$ have been proposed as QSL candidates [29-36]. Their crystal structure is close to the two-dimensional limit with the planar triangular lattice of the non-Kramers $Pr^{3+}$ ion [30,33-36]. The material hosts both static and dynamic structural disorder with the static mixture of Mg/Zn and Al on one of the Al sites necessary for charge balancing, and by the dynamic displacement of another Al site. A recent single crystal structure report also notes the presence of Pr positional disorder within its icosahedral oxygen cage [34,37].

Magnetization measurements on single crystals of $PrMgAl_{11}O_{19}$ revealed the proximity to the Ising limit [33-36]. Specific heat measurements excluded the formation of magnetic order down to 0.1 K [33-36]. The low-temperature specific heat in zero field follows a power-law behavior that can be fitted nearly to a $T^2$ dependence in the temperature range 0.4 to 2 K. However, above 0.5 T, a deviation from this power law is observed, which may indicate the formation of an energy gap [33-36]. In addition, powder inelastic scattering revealed an excitation at 1.5 meV, interpreted as an excitation between lowest lying CEF levels [33].

In this paper, we propose reconsidering the interpretation of the magnetic properties of $PrMgAl_{11}O_{19}$. By employing magnetic susceptibility, isothermal magnetization, and specific heat measurements on high-quality single crystals grown by the floating zone method, we confirm the proximity to Ising behavior and the absence of magnetic order. Based on our detailed study of the response of specific heat to the external magnetic field, we argue that the non-Kramers magnet $PrMgAl_{11}O_{19}$ is better described as a realization of an Ising triangular lattice antiferromagnet with an intrinsic transverse field, rather than as a QSL candidate. The TLAF structure and the non-Kramers nature of ions like $Pr^{3+}$ provide fertile ground for exploring the interplay between magnetic frustration and quantum fluctuations within the intrinsic transverse field model. In addition, we point out the big difference in magnetization and specific heat values between the different studies on $PrMgAl_{11}O_{19}$ single crystals, which indicates an unresolved dependence on growth conditions.

## Methods

The synthesis and subsequent crystal growth of $PrMgAl_{11}O_{19}$ were successfully carried out using a solid-state reaction and the optical floating zone (OFZ) method. Initial precursor binary oxides ($Pr_6O_{11}$, MgO, and $Al_2O_3$ of 99.99% purity) were calcined at 800°C for 24 hours to remove moisture or carbonate contamination. The oxides were then weighed keeping stoichiometric ratio, thoroughly mixed, and ground to ensure homogeneity of the mixture. The mixture was pressed into cylindrical rods with a 6 mm diameter and 100 mm length. Densification was achieved using a quasi-hydrostatic pressure of 2 tons for 15 minutes. The resulting rods were sintered in air at 1200°C for 72 hours to promote solid-state reaction and to improve the density of the material.

The initial attempt to grow $PrMgAl_{11}O_{19}$ crystals was conducted using a four-mirror optical floating zone furnace in a reducing environment to promote the formation of the $Pr^{3+}$ charge state. However, this approach encountered issues with the high evaporation of MgO, which hindered the successful growth of the crystals. In subsequent attempts, growth was performed under an air atmosphere with a slight overpressure of 1 atm and a flow rate of 3 L/min of air, solving the evaporation issue. The sintered rods were used both as feed and seed material for crystal growth. The feed and seed rods were counter-rotated during the growth at 30 rpm to improve temperature distribution and material mixing in the molten zone. The growth rate was maintained at 2 mm/h. Further details about the used synthesis method have been recently reported in our publication [38-39].

The growth resulted in a green ingot containing multiple grains with grain boundaries visible to the eye. The grains were separated using a wire saw and by breaking the ingot, and they were confirmed to be single crystals using backscattered Laue X-ray diffraction. Single-crystal X-ray diffraction (SCXRD) was performed at 95 K on a Rigaku SuperNova diffractometer equipped with an Atlas S2 CCD detector, using a mirror-collimated Mo Kα (λ = 0.71073 Å) radiation from a micro-focus sealed tube. Diffraction data integration, scaling, and adsorption correction were done using CrysAlis Pro [40] with an empirical absorption correction using spherical harmonics [41] combined with an analytical numeric absorption correction based on Gaussian integration over a multifaceted crystal model implemented in SCALE3 ABSPACK scaling algorithm. The structure was solved by charge flipping using the program Superflip [42] and refined by full-matrix least-squares on $F^2$ in Jana2020 [43].

To confirm that this solution from a small single crystal fragment represents the bulk, a larger single grain piece was crushed and checked by powder diffraction using a Panalytical Empyrean diffractometer with CuKα radiation in a capillary transmission parallel beam geometry at room temperature. Rietveld refinement confirmed the magnetoplumbite structure found by SCXRD, with room temperature lattice parameters $a$ = 5.58089(7) Å and $c$ = 21.9156(4) Å. X-ray fluorescence spectroscopy was used to confirm the compositional cation ratios of the prepared crystal, using an EDAX Orbis spectrometer with an Rh anode source (EKα = 20.216 keV) and polycapillary focusing optics.

The DC magnetic susceptibility was measured using a Quantum Design SQUID Magnetic Property Measurement System (MPMS) and a Vibrating Sample Magnetometer (VSM) on a Physical Property Measurement System (PPMS). The specific heat measurements were performed using a PPMS with the relaxation method. Below 2 K, the absence of magnetic field dependence of the addenda signal within experimental resolution was checked for each field of the measurement sequence. Specific heat measurements show similar results between samples grown with the two different atmospheres; however, single-crystal (directional) magnetization could be measured only on the larger crystals grown in air. Multiple pieces of single crystals from two growths (in air) were used for the reported bulk measurements with the results proving the sample quality and reproducibility of the response from this growth technique. This paper shows the results obtained from a piece of 29.7 mg. Single crystals of the non-magnetic analog $LaMgAl_{11}O_{19}$ were also grown in air using the floating zone method, and their specific heat measurement was used to estimate the phononic contribution to specific heat and therefore isolate the magnetic contribution ($C_m$) of the Pr-counterpart. Crystals of $LaMgAl_{11}O_{19}$, however, contained a tiny amount of magnetic impurities influencing the specific heat at low temperature; therefore, in the low-temperature limit (< 2 K), the phononic background was obtained from a $T^3$ extrapolation of the measured specific heat of $LaMgAl_{11}O_{19}$.

## Results

### Structural Characterization and CEF Calculations

Our single-crystal and powder diffraction data confirm the previous identification of $PrMgAl_{11}O_{19}$ as crystallizing in a magnetoplumbite structure [30,33-37,44-45]. This structure nominally consists of two-dimensional layers of $Pr^{3+}$ ions sitting within a 12-fold oxygen cage with $\bar{6}m2$ site symmetry. These polyhedra form a triangular lattice, sharing a single in-plane oxygen vertex with each neighboring $PrO_{12}$ polyhedron. The magnetic triangular layers are well separated, $c/2 \approx 11$ Å, by a spinel-like block of aluminum and magnesium oxide polyhedral (Fig. 1a).

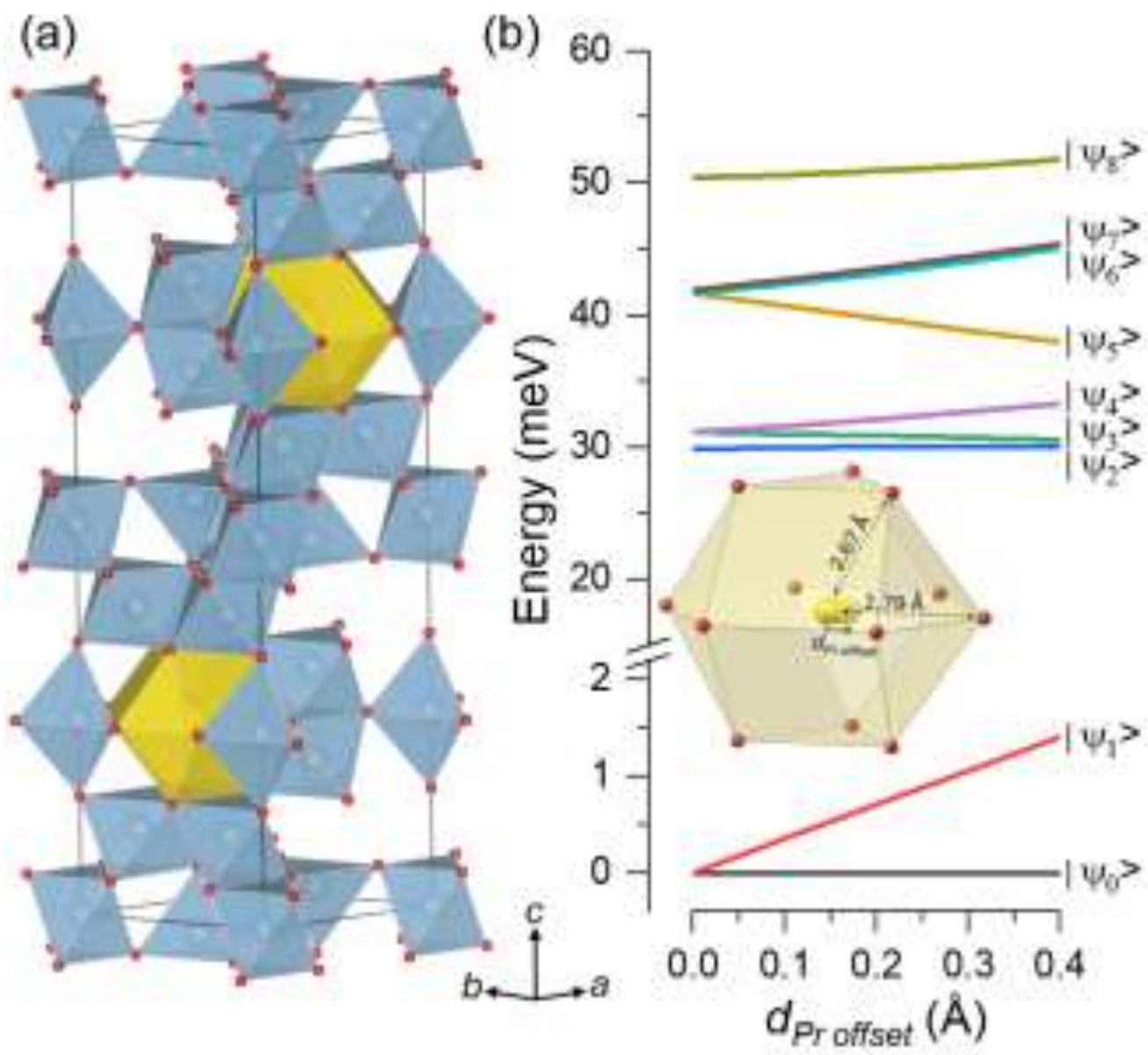

FIG. 1: The magnetoplumbite structure of $PrMgAl_{11}O_{19}$ (a), with $PrO_{12}$ polyhedra (yellow), forming a triangular lattice within *ab*-planes, separated by spinel-like blocks. The point-charge calculation of the crystal electric field split electronic energy levels (b), and the effect on the energies of these levels on displacing the $Pr^{3+}$ ion away from the center of the polyhedron.

Our single-crystal structure solution closely matches the structure solution previously reported [34, 37], so it will not be described in full here but is given in the supplementary information (SI). The structure actually hosts three separate sources of structural disorder: a site of randomly mixed Mg/Al occupation within the center of the spinel-like block, considered far enough from the magnetic lattice to only play a minor role; a disordered $Al^{3+}$ ion situated within a double-well potential of a triangular bipyramidal oxygen cage, between $PrO_{12}$ polyhedra in the triangular lattice plane; and the reported random positional disorder of the $Pr^{3+}$ ion within its polyhedron.

However, important to the discussion of the magnetic properties is the claim of $Pr^{3+}$ positional disorder in the structure, which was only noted in one of the several previous reports [34]. We find the best match between data and structural model with partial occupancy of $Pr^{3+}$ split between the high symmetry 2d Wyckoff site at the center of the $PrO_{12}$ polyhedron and an in-plane but off-center 6*h* site, with occupancies 0.837(8):0.016(3), respectively. This off-center site sits $d$ = 0.847 Å away from the polyhedron center and is evident when inspecting Fourier-difference electron density distribution maps with and without the split occupation (SI).

The two different sites of the $Pr^{3+}$ ion in the $PrO_{12}$ polyhedrons provide different crystal field environments, therefore we may expect different splitting of CEF levels for the two different sites. Conversely, $Pr^{3+}$ ions sitting at the off-center lower symmetry site will experience different crystalline electric fields that could further split the CEF energy levels. To test this, we performed point-charge calculations using the PyCrystalField software package to qualitatively inspect the effects of $Pr^{3+}$ positional disorder [60]. Eigenvalues and eigenvectors calculated based on only the high-symmetry central site of our solved crystal structure are comparable to those previously determined by point-charge calculations, refined using magnetization data [34]. Fig. 1(b) shows the resulting dependence of the calculated energy levels upon moving the Pr away from the center of the polyhedron and towards the disordered site found occupied in diffraction experiments. The previously calculated degeneracy of the ground state is broken, resulting in a lowest-lying non-magnetic singlet state and a low-lying excited state. With a displacement of just $d$ = 0.4 Å, a splitting of ~1.5 meV can be observed.

**Magnetization Measurements**

The temperature-dependent magnetic susceptibility of $PrMgAl_{11}O_{19}$ was measured with applied fields along and perpendicular to the *c*-axis in a magnetic field of 1 T as a function of temperature (Fig. 2a and 2b). The susceptibility in the field applied along the *c*-axis follows the Curie-Weiss (CW) law below 80 K. At higher temperatures, it deviates from CW behavior, likely due to the influence of excited CEF levels. A CW fitting in the 30-80 K temperature range yields a CW temperature of $\theta_{CW}$ = -11 K, in agreement with previous reports [30,33,34]. This value implies a magnetic interaction of $J = -2k_B\theta_{CW}/3$ = 0.63 meV. The CW fit gives the effective magnetic moment $\mu_{eff}$ = 4.3 μB. In contrast, the susceptibility measured with an applied field in the *ab*-plane (Fig. 2b) is significantly lower (approximately 20-fold at 2 K) compared to the *c*-axis direction. It does not follow the Curie-Weiss law, indicating significant contributions from excited CEF levels.

The isothermal magnetization of $PrMgAl_{11}O_{19}$ was measured with the magnetic field applied along the *c*-axis and within the *ab*-plane at temperatures between 2 K and 30 K, and in a field up to 14 T (Fig. 2c). The results confirm significant magnetic anisotropy in the crystal, with the magnetization along the *c*-axis being substantially higher than in the *ab*-plane. In the *ab*-plane, the magnetization curves are nearly linear within the measured field range, indicating a paramagnetic response without signs of saturation. In contrast, the magnetization measured with a field along the *c*-axis deviates from linearity in a relatively small field. Despite this deviation, the magnetization does not saturate up to 14 T. This anisotropic behavior indicates that $PrMgAl_{11}O_{19}$ is an effective Ising magnet, where the magnetic moments preferentially align along the *c*-axis, perpendicular to the well-separated triangular lattice planes. The absence of a magnetization plateau in $PrMgAl_{11}O_{19}$ contrasts with theoretical predictions for $S_{eff} = 1/2$ TLAF [61]. Therefore, it hints at the splitting of the lowest energy quasi-doublet into two singlets and for a pseudo-spin S = 1/2 Hamiltonian [7].

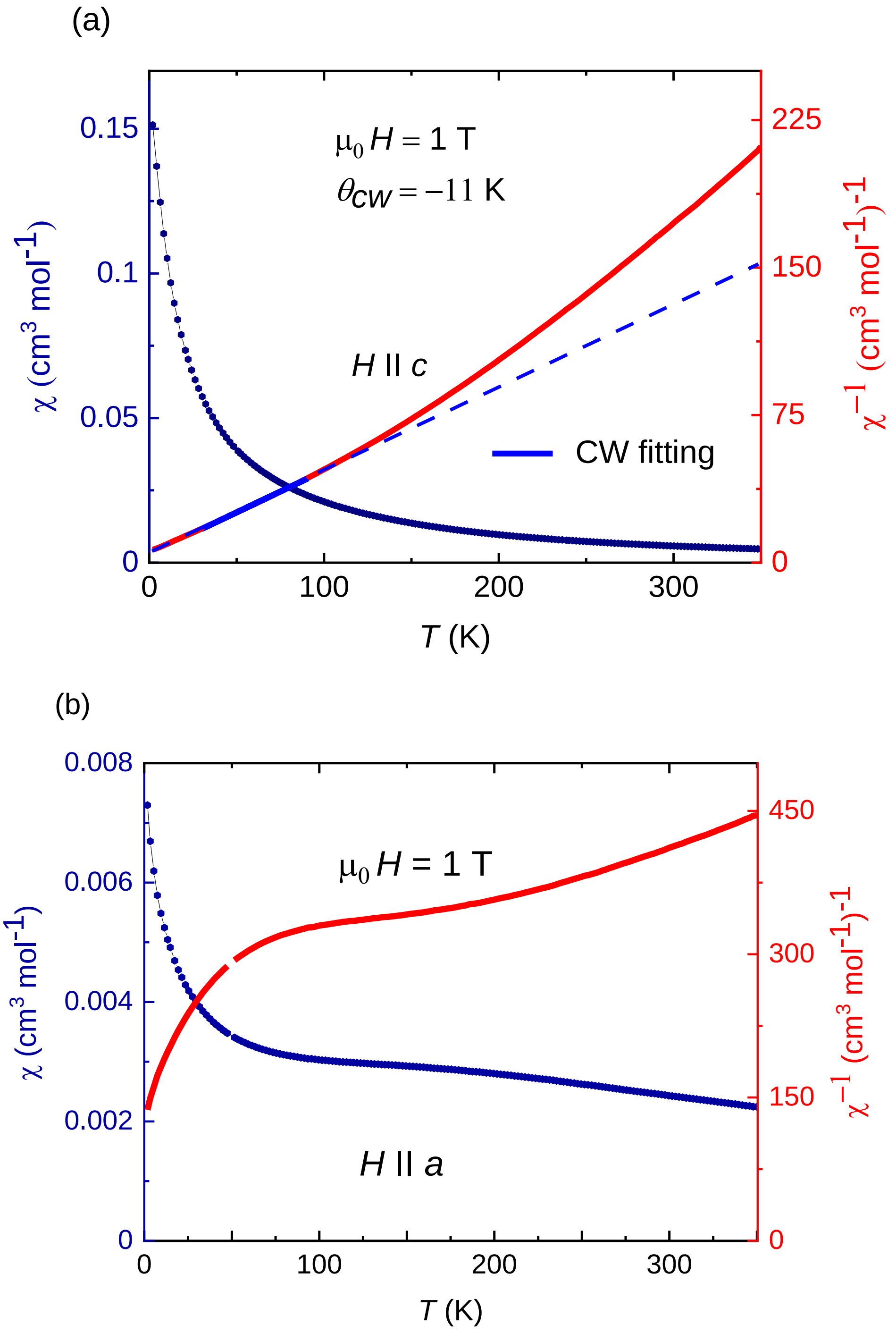

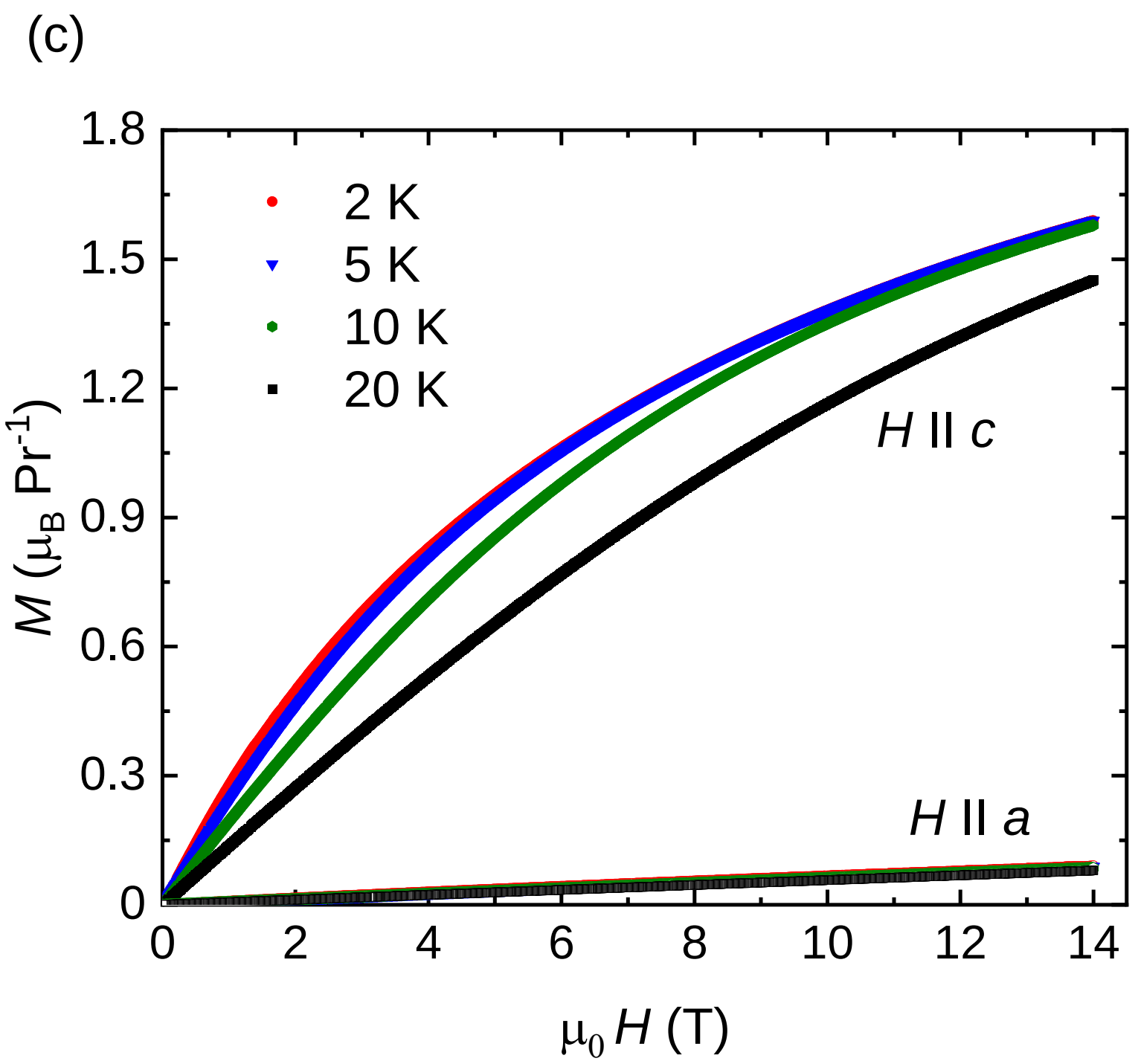

FIG. 2: Magnetic susceptibility and inverse susceptibility measured with an external magnetic field $\mu_0 H$ = 1 T, when *H* is applied parallel to the *c*-axis (a), and perpendicular to the *c*-axis (b). The blue line in (a) represents the Curie-Weiss (CW) fitting of the inverse susceptibility in the temperature range of 30-80 K, yielding a Curie-Weiss temperature ($\theta_{CW}$) of -11 K. (c) Isothermal magnetization as a function of the field at different temperatures between 2 K and 30 K for fields applied in the *ab*-plane and along the *c*-axis, demonstrating significant magnetic anisotropy in the system.

We highlight an important non-reproducibility of the magnetization of $PrMgAl_{11}O_{19}$ grown by different groups. All single crystals we investigated (two ingots grown in air and several single-grain pieces separated from each) reveal comparable magnetization values. In contrast, independent studies on several crystals synthesized using similar floating zone methods reported higher magnetization values. As an example, the magnetization measured at 2 K with a magnetic field applied along the *c*-axis reaches a value of 1.2 μB/Pr (our work) at 7 T, while values of 1.8 μB/Pr [33,34], 1.5 μB/Pr [35], and 1.6 μB/Pr [36] have also been reported.

**Specific Heat**

The specific heat of $PrMgAl_{11}O_{19}$ and its non-magnetic analog $LaMgAl_{11}O_{19}$ are represented in Fig. 3. The absence of any sharp anomaly down to 0.4 K indicates that no long-range magnetic order is present within this temperature range. $C_m/T$ vs. T curve at zero field shows a broad hump with a maximum at 5 K and a shoulder at 13 K (Fig. 3b). The total entropy reaches Rln(2) approaching 80 K, confirming that two levels exhaust the low-energy degrees of freedom.

To test whether this double hump could be the signature of the depopulation upon cooling of one of the two singlets of the non-Kramers quasi-doublet of $Pr^{3+}$, we have compared the specific heat with a Schottky model. We propose that the double hump could come from two slightly different crystallographic sites for $Pr^{3+}$ revealed by single-crystal XRD [SI]. The two different Pr sites would be associated with different energy splitting of the quasi-doublet due to different local CEF. Therefore, we have fitted the data to a two-gap Schottky model [SI] of the form:

$$\frac{C_m}{T}(T) = \frac{R}{T}\sum_{j=1}^{2} f_j \left(\frac{\Delta_j}{k_B T}\right)^2 \frac{Exp\left(\frac{\Delta_j}{k_B T}\right)}{\left(1 + Exp\left(\frac{\Delta_j}{k_B T}\right)\right)^2}$$

where $f_j$ is the fraction of Pr ions contributing to the $j$-th Schottky term, and $\Delta_j$ is the corresponding energy gap.

At zero field, the two-gap Schottky model yields a smaller gap $\Delta_1 \approx 1.26$ meV, corresponding to the first hump near 5 K, and a larger gap $\Delta_2 \approx 5.16$ meV, associated with the shoulder at 13 K. The deviation from this simple two-gap fit is larger at zero field but still captures the main double-hump structure. The zero-field humps are broader than the Schottky model predicts, suggesting a distribution of $Pr^{3+}$ local environments and/or the effects of spin-spin correlations that are not captured by our simplified two-site model; nevertheless, the model suitably captures the main features. The observation of Rln(2) up to ~80 K confirms that the low-energy quasi-doublet is isolated from higher CEF levels, suggesting that the excitation observed at ~1.5 meV in inelastic neutron scattering (INS) [33] arises from the splitting between the two singlets of the quasi-doublet rather than from magnetic interactions. In contrast to point-charge calculations, which predict a splitting of the two lowest-energy singlets only for $Pr^{3+}$ ions displaced from the center of the oxygen cage, the specific heat measurements indicate that the quasi-doublet is split for all $Pr^{3+}$ ions, with a larger splitting for a fraction of them—presumably those in the off-center secondary position. This discrepancy highlights the approximate nature of point-charge calculations, which, as noted in our and previous studies [34], fail to accurately predict the splitting of the quasi-doublet for $Pr^{3+}$ ions at the high-symmetry central site due to oversimplified assumptions about the local crystal CEF.

Under an applied magnetic field, the smaller gap $\Delta_1$ undergoes a strong Zeeman shift, moving the Schottky peak to higher temperatures with increasing field strength, while the larger gap $\Delta_2$ is less sensitive, displaying only a minor shift. Below 7 T, the two hump-like features become more separated and discernible. The field dependence of $\Delta_1$ and $\Delta_2$ follows the theoretical model proposed in Ref.[22] for quantum-induced magnets as shown in the SI. This approach gives the approximative $g$-factor values 5.7 and 9.8 for $\Delta_1$ and $\Delta_2$, respectively. Assuming a pseudospin $S_{\mathrm{eff}} = 1/2$ description, the CW analysis of susceptibility yields an effective $g$-factor of $g_c = 5.0$, in close agreement with previous Electron Spin Resonance (ESR) measurements reporting $g_c = 5.1$ [34].

At high fields ($\mu_0 H$ = 7 T and 9 T), the two-gap Schottky model reproduces the experimental data well, with fractions $f_1 \approx 0.42$ and $f_2 \approx 0.58$ reflecting the contributions of the two $Pr^{3+}$ environments. We cannot fully reconcile the different populations of the two sites between the SCXRD structure solution at 95 K (~84:16) and this low-temperature Schottky fit (~58:42), but hypothesize that the population may be temperature-dependent resulting in a higher proportion of off-center $Pr^{3+}$ ions at low-temperature. Temperature-dependent diffraction studies would be needed to confirm this.

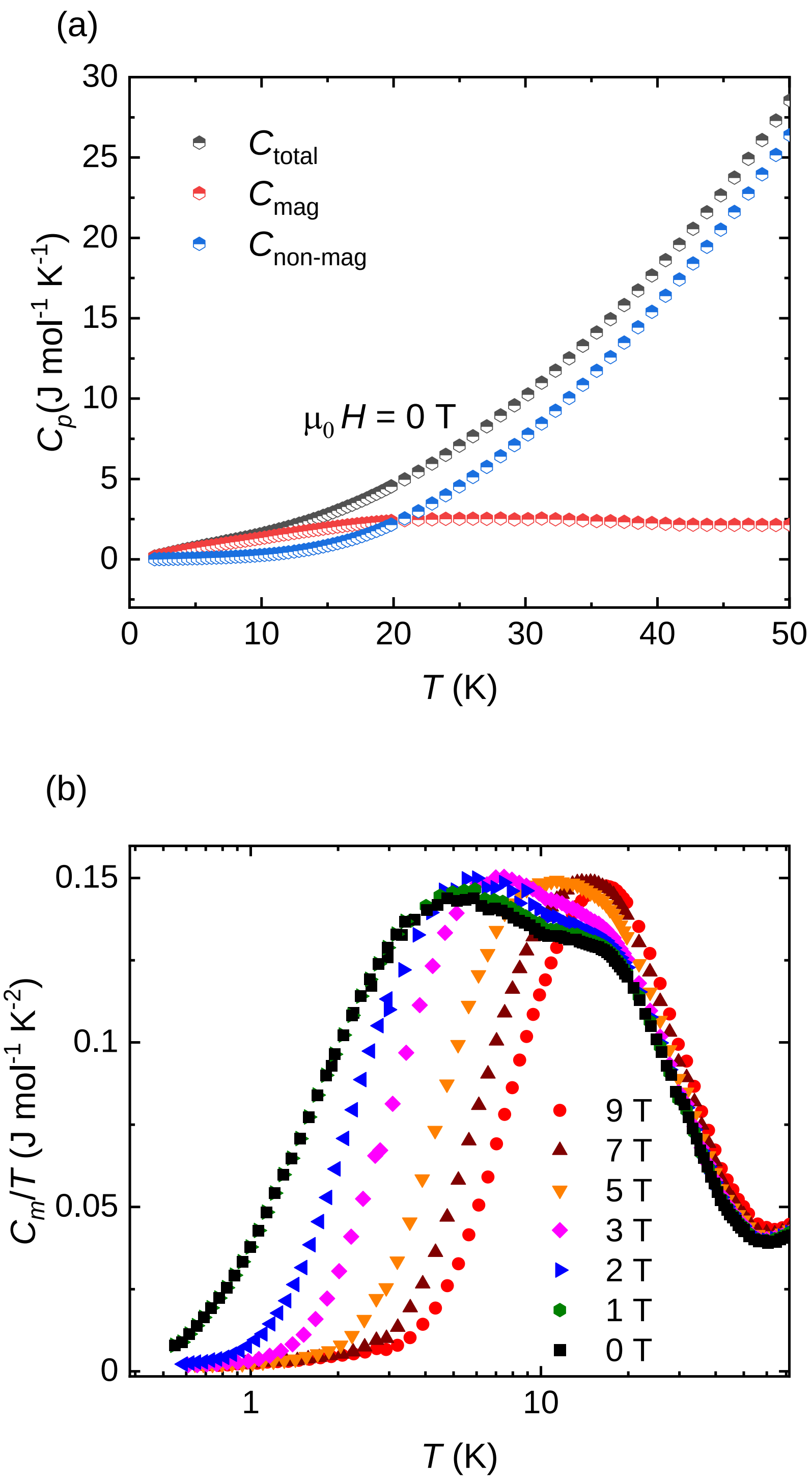

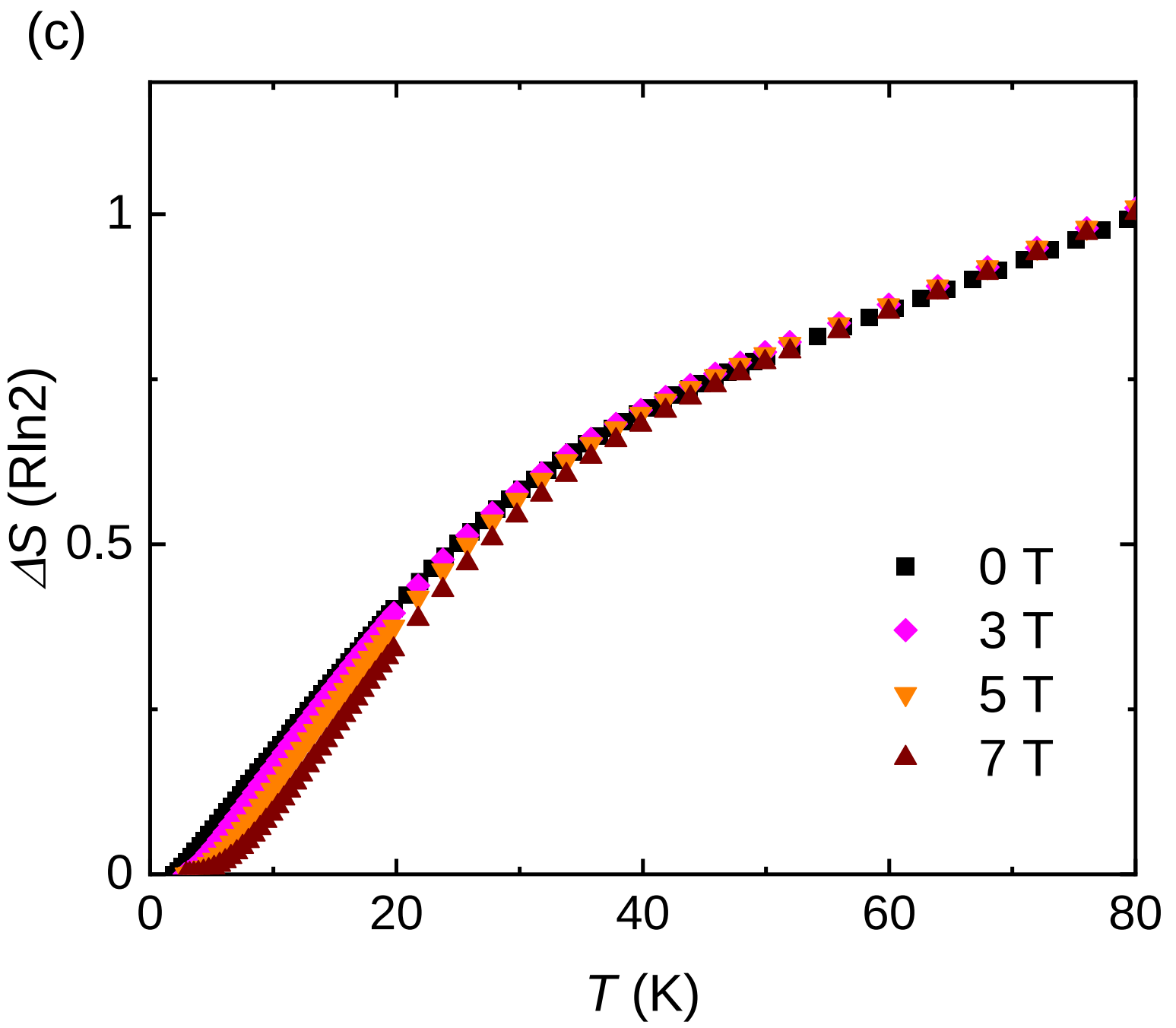

FIG. 3: Specific heat measurements for $PrMgAl_{11}O_{19}$. (a) Total specific heat ($C_p$) as a function of temperature, compared with the non-magnetic analog $LaMgAl_{11}O_{19}$ ($C_{non\text{-}m}$) to isolate the magnetic contribution ($C_m$). (b) $C_m/T$ as a function of $T$, showing a broad hump that shifts to higher temperatures with an increasing applied magnetic field. (c) Magnetic entropy change ($\Delta S$) as a function of temperature, approaching Rln2 around 80 K.

To further investigate the response of the specific heat under an external magnetic field, we measured the low-temperature specific heat with small magnetic fields applied along the easy magnetization axis *c* (Fig. 5a). At very low fields, up to $\mu_0 H = 0.25$ T, the specific heat exhibits a non-monotonic behavior, with an initial increase in $C_m$ as the field strength rises. This enhancement suggests that weak magnetic fields may enhance certain low-energy excitations or interactions before other effects dominate. However, above $\mu_0 H = 0.25$ T, $C_m$ begins to decrease with increasing field, indicating a suppression of these contributions as the field strength grows.

At low temperatures ($k_B T \ll \Delta$), the specific heat deviates from the Schottky model and follows a power-law behavior $C_m = cT^{\alpha}$, with $\alpha = 1.9$ (Fig. 5b), indicating the presence of gapless interactions. However, this behavior changes with the application of higher magnetic fields. Above $\mu_0 H = 0.5$ T, the power-law behavior is no longer observed. One possibility is that the specific heat comprises both gapped and gapless components, though the current data are insufficient to confidently fit such a composite model. Interestingly, at $\mu_0 H = 9$ T, the gapless behavior re-emerges, further highlighting the intricate field-dependent response of the system.

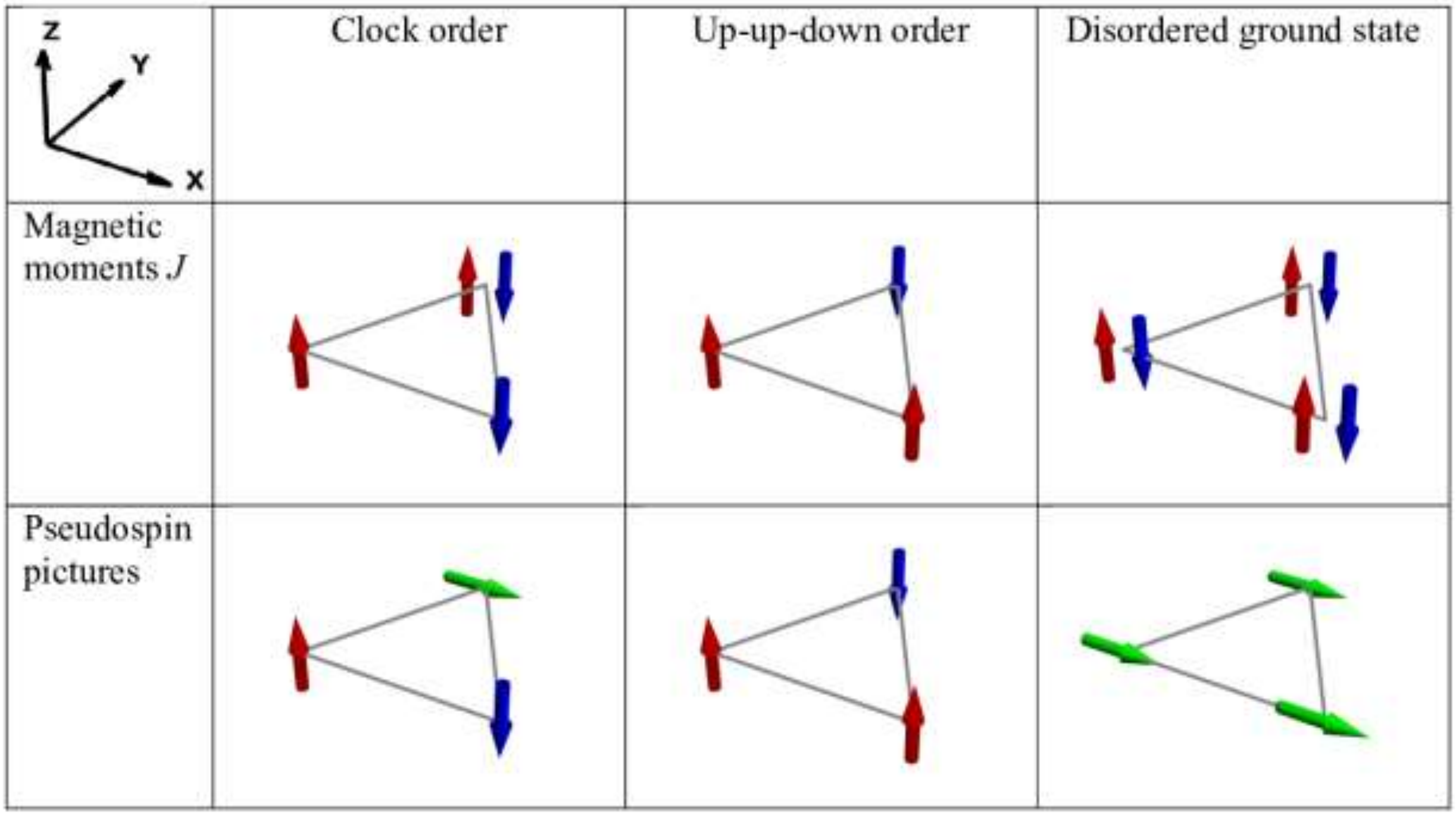

FIG. 4: Schematic view of magnetic phases in competition in triangular magnets with non-Kramers magnetic ions such as $Pr^{3+}$ or $Tm^{3+}$. The superposition |0> = |$J_z$ = + 4 > + |$J_z$ = - 4 > is represented by both a moment up and a moment down or in the pseudospin picture by a pseudospin pointing towards the transverse direction *x*. These three phases were located on the *h* – *H* phase diagram based on previous theoretical work [7].

## Discussions

The integrated entropy of Rln2 and the two broad Schottky-like features in $C_m/T$ confirm a two-level description of all $Pr^{3+}$ ions within the structure. Indeed, the extracted effective g-factor (from both CW and Schottky analysis) is close to the upper limit of $2Jg_J \approx 6.42$ for J = 4 and $g_J = 0.8$. This quantitatively supports the interpretation that the ground state is predominantly |$m_J$ = ±4> based [20, 46]. Similarly to other studies in Pr-based magnets [10, 22], we propose a description of the $Pr^{3+}$ ion low-temperature magnetism in $PrMgAl_{11}O_{19}$ as the two-level system:

|0> = |Jz = + 4 > + |Jz = - 4 >

|1> = |Jz = + 4 > - |Jz = - 4 > (1)

When the energy splitting between the two singlets, Δ, is comparable to the magnetic interaction, J, the magnetism can be described by an effective Hamiltonian on pseudospins $S_{\mathrm{eff}}$ = 1/2 [5-7, 10, 22, 47]. The energy splitting Δ is equivalent to an intrinsic transverse field acting on the pseudospin component $S^x$, and the Hamiltonian can be written as:

$$H = \sum_{<i,j>} J S_i^z S_j^z - \sum_i h S_i^x - \sum_i \mu_0 \mu_B g_c H S_i^z. \quad (2)$$

We propose that this entropy change fully corresponds to the population of the excited singlet |1>, as seen in $Pr_2NiTiO_6$, $TmMgGaO_4$, and $Pr_3WBO_9$ [5, 6, 10], rather than including the formation of correlations between magnetic moments, which has been recently proposed [33-36]. This scenario is further confirmed by the previous observation of a Q-independent excitation just below 1.5 meV (close to our Schottky analysis) by inelastic neutron scattering on a powder sample [33]. As $C_m$ follows a power-law behavior between 0.5 K < 2 K with $\alpha = 1.9$, this nearly quadratic behavior excludes the formation of an excitation gap of the order of the intrinsic transverse field h ~ 1.26 meV. Such quadratic behavior was observed in other non-Kramers magnets such as $Pr_3WBO_9$ [10] and $KTmSe_2$ [11], and it was explained as a consequence of the presence of an antiferromagnetic interaction $J$ comparable with the splitting $h$. Indeed, at T = 0, only the lowest singlet |0> state is populated. Upon increasing temperature, the higher singlet |1> state starts gradually to be populated. The antiferromagnetic interaction $J$ favors the population of the highest singlet |1> on neighboring sites, such that magnetic moments arise, promoting antiferromagnetic correlations. This phenomenon leads to a power-law behavior of $C_m(T)$ [10] and is referred to as induced quantum magnetism [22].

The non-monotonous behavior explains that the antiferromagnetic interactions favor induced moments with antiferromagnetic correlations, and the external magnetic field favors induced moments along the direction of the field. Both can be satisfied by the formation of short-range up-up-down correlations. Therefore, the increase in specific heat with an applied magnetic field up to $\mu_0 H$ = 0.25 T could be explained by temperature-induced magnetic moments with up-up-down correlations. However, under the application of a high magnetic field (5 T - 9 T), the magnetic contribution to the specific heat is not fully suppressed, and a nearly field-independent component remains. This was previously observed in other studies, which failed to explain it within the assumption that $PrMgAl_{11}O_{19}$ is a QSL [35, 36]. Considering the specificity of the $Pr^{3+}$ non-Kramers ion, we propose that the magnetic contribution to the low-temperature specific heat could be the sum of a contribution from the induced dipolar moment and a contribution from the multipolar degree of freedom of the singlet ground state [19, 20, 48]. While the external magnetic field suppresses the contribution from dipolar moments, the contribution from the multipolar degree of freedom would remain. In this context, a question remains for future theoretical studies, whether the apparent $T^2$ behavior of low-temperature specific heat in a high magnetic field of $\mu_0 H$ = 9 T is compatible with this scenario.

Theoretical studies of the quantum Ising magnetism on a triangular lattice predict the competition between three different magnetic states: the clock order, the up-up-down order, and the disordered state [6,7,9]. Representations of these phases' magnetic moment configurations, as well as their pseudospin representations, are shown in Fig. 4. These phases have previously been identified on the transverse field - external field ($h$-$H$) phase diagram by quantum Monte Carlo calculations [7]. In the absence of an external field, the magnetic quantum critical point between the clock order and the disordered state is located at $h_c/J$ ~ 0.8 [7,9]. With h ~ 1.26 meV and J ~ 0.6 meV, $PrMgAl_{11}O_{19}$ is located around $h/J$ ~ 2, which is above the critical value, in the regime where magnetic order is not expected.

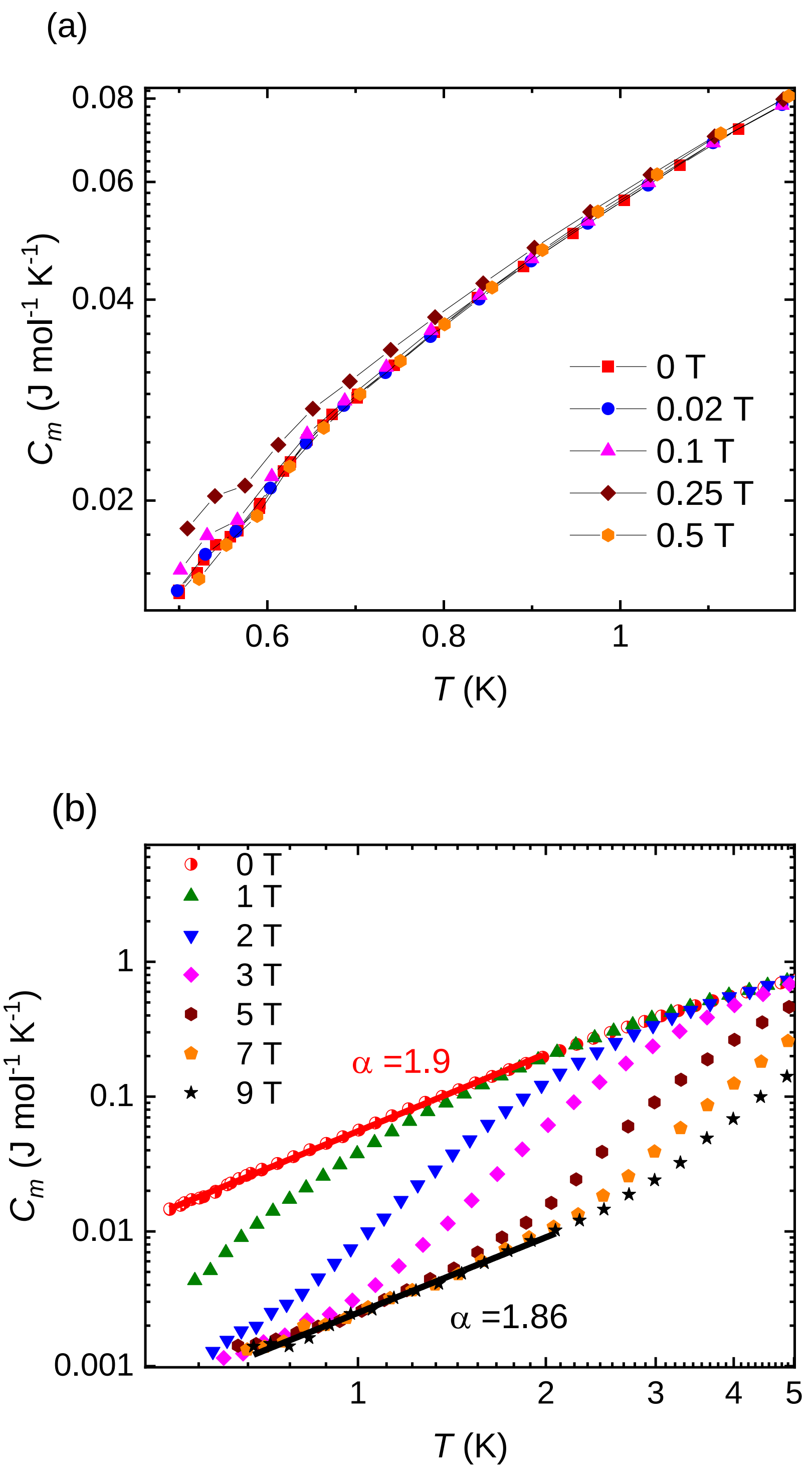

FIG. 5: Magnetic contribution $C_m$ to the specific heat in or $PrMgAl_{11}O_{19}$ as a function of temperature under a magnetic field up to 0.5 T in (a) and up to 9T in (b). The solid lines indicate power law fitting.

Numerous $Pr^{3+}$ and $Tm^{3+}$ based triangular magnets such as $CsPrSe_2$ [49], $PrZn_3P_3$ [50], $PrTa_7O_{19}$ [51], $Pr_2O_2CO_3$ [52], and $NaBaTm(BO_3)_2$ [53] show Van Vleck paramagnetism and suppression of the magnetic contribution to the specific heat at low temperature. These features indicate that the splitting of the quasi-doublet strongly overcomes the magnetic interactions. On the contrary, $PrMgAl_{11}O_{19}$ with $h/J$ ~ 2 is much closer to the quantum critical point at $h_c/J = 0.8$ than these Van Vleck paramagnets and even closer than other model compounds, e.g., $KTmSe_2$ with $h/J$ ~ 3.9 [11]. In addition, $PrMgAl_{11}O_{19}$ shows a large difference between the value of the intrinsic field $h$ and the first CEF excitation Δ, as shown qualitatively by our specific heat and quantitatively from the point charge calculation. As a consequence, the interplay between the intrinsic transverse field and the CEF excitations can be neglected in $PrMgAl_{11}O_{19}$, in contrast to $NaTmSe_2$ [54], $KTmSe_2$ [11], and $PrZnAl_{11}O_{19}$ [32]. Thus, $PrMgAl_{11}O_{19}$ is a promising compound for the study of disorder-induced quantum magnetism in the vicinity of a quantum critical point. The absence of 1/3 magnetic plateau further supports this description. We suggest that the search for a QSL induced by the proximity of the Ising limit should rather be focused on Kramers magnets with effective spin $S_{\mathrm{eff}} = 1/2$, such as $NdTa_7O_{19}$ [55].

We also pointed out a strong sample dependence between the different single-crystal studies of the magnetic properties of $PrMgAl_{11}O_{19}$ [33-36]. All these studies were based on crystals grown by the floating zone method. The main difference between the growths seems to be the choice of the atmosphere between air in this work and [33,36], and Ar in [34,35]. However, there is no apparent correlation between the choice of the atmosphere and magnetic properties. The sample dependence might come from $Pr^{3+}$ vacancies, oxygen vacancies, or the presence of $Pr^{4+}$ impurities. Detailed studies of the magnetic properties of $PrMgAl_{11}O_{19}$ depending on growth conditions are desirable.

## Conclusions

In conclusion, we successfully grew single crystals of the QSL candidate $PrMgAl_{11}O_{19}$ with the floating zone method and performed SCXRD, magnetization, and specific heat measurements on them. Based on our results and point-charge calculations of the crystal electric field Hamiltonian, we describe the magnetism of $PrMgAl_{11}O_{19}$ as quantum Ising magnetism that is well captured by the intrinsic transverse field model, not the previously proposed QSL state. We claim that $PrMgAl_{11}O_{19}$ is a promising compound for the study of quantum Ising magnetism on a triangular lattice. It lies in the regime of quantum-disordered magnetism, where the intrinsic transverse field is strong enough to prevent the formation of magnetic order yet weak enough to promote the formation of magnetic correlations. The magnetic ground state of $PrMgAl_{11}O_{19}$ deserves further investigation. In particular, single-crystal inelastic neutron scattering would be valuable to study the magnetic excitations of its exotic magnetic ground-state phase.

## Acknowledgment

We acknowledge funding from the Charles University in Prague within the Primus research program with grant number PRIMUS/22/SCI/016, the Czech Ministry of Education Youth and Sports (grant number LUABA24056), and the Grant Agency of Univerzita Karlova (grant number 438425). Crystal growth, structural analysis, and magnetic properties measurements were carried out in the MGML (http://mgml.eu/), supported within the Czech Research Infrastructures program (project no. LM2023065). We acknowledge Martin Míšek and Martin Žáček for their technical help with magnetization measurements with VSM technique.

## References

[1] W. Liu et al., "Rare-Earth Chalcogenides: A Large Family of Triangular Lattice Spin Liquid Candidates," *Chin. Phys. Lett.* **35**, 117501, November 2018, https://doi.org/10.1088/0256-307X/35/11/117501

[2] L. Savary and L. Balents, "Quantum spin liquids: a review," *Rep. Prog. Phys.* **80**, 016502, December 2016, https://doi.org/10.1088/0034-4885/80/1/016502

[3] Y. Li et al., "Spin liquids in geometrically perfect triangular antiferromagnets," *J. Phys.: Condens. Matter* **32**, 224004, May 2020, https://doi.org/10.1088/1361-648X/ab73e5

[4] T. Lancaster et al., "Quantum spin liquid," *Contemp. Phys.* **64**, 129, December 2023, https://doi.org/10.1080/00107514.2023.2284522

[5] Y. Li et al., "Gapped ground state in the zigzag pseudospin-1/2 quantum antiferromagnetic chain compound $PrTiNbO_6$," *Phys. Rev. B* **97**, 184434, May 2018, https://doi.org/10.1103/PhysRevB.97.184434

[6] Y. Li et al., "Partial Up-Up-Down Order with the Continuously Distributed Order Parameter in the Triangular Antiferromagnet $TmMgGaO_4$," *Phys. Rev. X* **10**, 011007, January 2020, https://doi.org/10.1103/PhysRevX.10.011007

[7] C. Liu et al., "Intrinsic quantum Ising model on a triangular lattice magnet $TmMgGaO_4$," *Phys. Rev. Res.* **2**, 043013, October 2020, https://doi.org/10.1103/PhysRevResearch.2.043013

[8] Y. Jiang and T. Emig, "String Picture for a Model of Frustrated Quantum Magnets and Dimers," *Phys. Rev. Lett.* **94**, 110604, March 2005, https://doi.org/10.1103/PhysRevLett.94.110604

[9] Z. Zhou et al., "Quantum dynamics of topological strings in a frustrated Ising antiferromagnet," *npj Quantum Mater.* **7**, 60, June 2022, https://doi.org/10.1038/s41535-022-00471-8

[10] J. Nagl et al., "Excitation spectrum and spin Hamiltonian of the frustrated quantum Ising magnet $Pr_3BWO_9$," *Phys. Rev. Res.* **6**, 023267, April 2024, https://doi.org/10.1103/PhysRevResearch.6.023267

[11] S. Zheng et al., "Exchange-renormalized crystal field excitations in the quantum Ising

magnet $KTmSe_2$," *Phys. Rev. B* **108**, 054435, August 2023, https://doi.org/10.1103/PhysRevB.108.054435

[12] J. Chaloupka, "Emergent transverse-field Ising model in $d^4$ spin-orbit Mott insulators," *Phys. Rev. B* **109**, L020403, January 2024, https://doi.org/10.1103/PhysRevB.109.L020403

[13] M. Moreno-Cardoner et al., "Case study of the uniaxial anisotropic spin-1 bilinear-biquadratic Heisenberg model on a triangular lattice," *Phys. Rev. B* **90**, 144409, October 2014, https://doi.org/10.1103/PhysRevB.90.144409

[14] U. F. P. Seifert and L. Savary, "Phase diagrams and excitations of anisotropic S=1 quantum magnets on the triangular lattice," *Phys. Rev. B* **106**, 195147, November 2022, https://doi.org/10.1103/PhysRevB.106.195147

[15] S. Hayami and K. Hattori, "Multiple-q Dipole–Quadrupole Instability in Spin-1 Triangular-lattice Systems," *J. Phys. Soc. Jpn.* **92**, 124709, December 2023, https://doi.org/10.7566/JPSJ.92.124709

[16] K. Hattori and H. Tsunetsugu, "Antiferro quadrupole orders in non-Kramers doublet systems," *J. Phys. Soc. Jpn.* **83**, 034709, March 2014, https://doi.org/10.7566/JPSJ.83.034709

[17] T. Yanagisawa et al., "Quadrupolar susceptibility and magnetic phase diagram of $PrNi_2Cd_{20}$ with non-Kramers doublet ground state," *Philos. Mag.* **100**, 1268, May 2020, https://doi.org/10.1080/14786435.2020.1711505

[18] T. Onimaru and H. Kusunose, "Exotic quadrupolar phenomena in non-Kramers doublet systems - The cases of $PrT_2Zn_{20}$ (T = Ir, Rh) and $PrT_2Al_{20}$ (T = V, Ti)," *J. Phys. Soc. Jpn.* **85**, 082002, August 2016, https://doi.org/10.7566/JPSJ.85.082002

[19] C. Liu et al., "Selective measurements of intertwined multipolar orders: Non-Kramers doublets on a triangular lattice," *Phys. Rev. B* **98**, 045119, July 2018, https://doi.org/10.1103/PhysRevB.98.045119

[20] Y. Shen et al., "Intertwined dipolar and multipolar order in the triangular-lattice magnet $TmMgGaO_4$," *Nat. Commun.* **10**, 4530, October 2019, https://doi.org/10.1038/s41467-019-12433-3

[21] N. Tang et al., "Spin–orbital liquid state and liquid–gas metamagnetic transition on a pyrochlore lattice," *Nat. Phys.* **19**, 92, January 2023, https://doi.org/10.1038/s41567-022-01801-5

[22] P. Thalmeier and A. Akbari, "Induced quantum magnetism in crystalline electric field singlet ground state models: Thermodynamics and excitations," *Phys. Rev. B* **109**, 115110,

March 2024, https://doi.org/10.1103/PhysRevB.109.115110

[23] B. Grover, "Dynamical properties of induced-moment systems," *Phys. Rev.* **140**, A1944, December 1965, https://doi.org/10.1103/PhysRev.140.A1944

[24] R. J. Birgeneau et al., "Physical Review Letters 29," *Phys. Rev. Lett.* **27**, 1530, November 1971, https://doi.org/10.1103/PhysRevLett.27.1530

[25] D. B. McWhan et al., "Neutron scattering study of pressure-induced antiferromagnetism in PrSb," *Phys. Rev. B* **20**, 3442, November 1979, https://doi.org/10.1103/PhysRevB.20.3442

[26] R. Moessner and S. L. Sondhi, "Ising models of quantum frustration," *Phys. Rev. B* **63**, 224401, May 2001, https://doi.org/10.1103/PhysRevB.63.224401

[27] Z. Hu et al., "Evidence of the Berezinskii-Kosterlitz-Thouless phase in a frustrated magnet," *Nat. Commun.* **11**, 5631, November 2020, https://doi.org/10.1038/s41467-020-19380-6

[28] H. Li et al., "Kosterlitz-Thouless melting of magnetic order in the triangular quantum Ising material $TmMgGaO_4$," *Nat. Commun.* **11**, 1111, March 2020, https://doi.org/10.1038/s41467-020-14907-8

[29] M. Ashtar et al., "$REZnAl_{11}O_{19}$ (RE = Pr, Nd, Sm-Tb): A new family of ideal 2D triangular lattice frustrated magnets," *J. Mater. Chem. C* **7**, 10073, August 2019, https://doi.org/10.1039/C9TC02606A

[30] M. Ashtar et al., "Synthesis, structure and magnetic properties of rare-earth $REMgAl_{11}O_{19}$ (RE = Pr, Nd) compounds with two-dimensional triangular lattice," *J. Alloys Compd.* **802**, 146, September 2019, https://doi.org/10.1016/j.jallcom.2019.06.121

[31] D. Ni and R. J. Cava, "Ferrites without iron as potential quantum materials," *Prog. Solid State Chem.* **66**, 100346, December 2022, https://doi.org/10.1016/j.progsolidstchem.2022.100346

[32] H. Bu et al., "Gapless triangular-lattice spin-liquid candidate $PrZnAl_{11}O_{19}$," *Phys. Rev. B* **106**, 134428, October 2022, https://doi.org/10.1103/PhysRevB.106.134428

[33] Z. Ma et al., "Possible gapless quantum spin liquid behavior in the triangular-lattice Ising antiferromagnet $PrMgAl_{11}O_{19}$," *Phys. Rev. B* **109**, 165143, April 2024, https://doi.org/10.1103/PhysRevB.109.165143

[34] Y. Cao et al., "Synthesis, disorder and Ising anisotropy in a new spin liquid candidate $PrMgAl_{11}O_{19}$," *Mater. Futures* **3**, 035201, September 2024, https://doi.org/10.1088/2752-5724/ad4c5e

[35] N. Li et al., "Ising-type quantum spin liquid state in $PrMgAl_{11}O_{19}$," *Phys. Rev. B* **110**, 134401, *Oct 2024*, https://doi.org/10.1103/PhysRevB.110.134401

[36] C. Tu et al., "Gapped quantum spin liquid in a triangular-lattice Ising-type antiferromagnet $PrMgAl_{11}O_{19}$," *Phys. Rev. Research* **6**, 043147, Nov. 2024, https://doi.org/10.1103/PhysRevResearch.6.043147

[37] S. C. Abrahams et al., "Laser and phosphor host $La_{1-x}MgAl_{11+x}O_{19}$ (x=0.050): Crystal structure at 295 K," *J. Chem. Phys.* **86**, 4221, April 1987, https://doi.org/10.1063/1.451883

[38] D. Staško et al., "The synthesis of the rare earth $A_2Zr_2O_7$ single crystals by simplified laser-heated floating hot zone and pedestal methods," *Mater. Today Chem.* **39**, 102153, July 2024, https://doi.org/10.1016/j.mtchem.2024.102153

[39] M. Klicpera et al., "Magnetic frustration in rare-earth zirconates $A_2Zr_2O_7$, the case of laser heated pedestal method synthesised A = Er, Tm, Yb, and Lu single crystals," *J. Alloys Compd.* **978**, 173440, March 2024, https://doi.org/10.1016/j.jallcom.2023.173440

[40] *CrysAlisPro*, Oxford Diffraction, Agilent Technologies UK Ltd, England, doi:10.1107/S160057672001554X

[41] R. C. Clark and J. S. Reid, "The analytical calculation of absorption in multifaceted crystals," *Acta Crystallogr. A* **51**, 887, November 1995, https://doi.org/10.1107/S0108767395007356

[42] L. Palatinus and G. Chapuis, "SUPERFLIP - A computer program for the solution of crystal structures by charge flipping in arbitrary dimensions," *J. Appl. Crystallogr.* **40**, 786, August 2007, https://doi.org/10.1107/S0021889807029232

[43] V. Petříček et al., "Jana2020 – a new version of the crystallographic computing system Jana," *Z. Kristallogr. - Cryst. Mater.* **238**, 271, July 2023, https://doi.org/10.1515/zkri-2023-0007

[44] D. Saber and A.-M. Lejus, "Elaboration and characterization of lanthanide aluminate single crystals with the formula $LnMgAl_{11}O_{19}$," *Mater. Res. Bull.* **16**, 1325, October 1981, https://doi.org/10.1016/0025-5408(81)90050-8

[45] M. Gasperin et al., "Influence of $M^{2+}$ ions substitution on the structure of lanthanum hexaaluminates with magnetoplumbite structure," *J. Solid State Chem.* **54**, 61, August 1984, https://doi.org/10.1016/0022-4596(84)90188-5

[46] Y. Li et al., "Absence of zero-point entropy in a triangular Ising antiferromagnet," *arXiv:2501.17052*, January 2025, https://doi.org/10.48550/arXiv.2501.17052

[47] G. Chen, "Intrinsic transverse field in frustrated quantum Ising magnets: Physical origin and quantum effects," *Phys. Rev. Res.* **1**, 033141, December 2019, https://doi.org/10.1103/PhysRevResearch.1.033141

[48] Y. Qin et al., "Field-tuned quantum effects in a triangular-lattice Ising magnet," *Sci. Bull.* **67**, 38, January 2022, https://doi.org/10.1016/j.scib.2021.09.006

[49] J. Xing et al., "Crystal synthesis and frustrated magnetism in triangular lattice $CsRESe_2$ (RE= La–Lu): Quantum spin liquid candidates $CsCeSe_2$ and $CsYbSe_2$," *ACS Mater. Lett.* **2**, 71, January 2019, https://doi.org/10.1021/acsmaterialslett.9b00003

[50] N. Kabeya et al., "Competing Exchange Interactions in Lanthanide Triangular Lattice Compounds $LnZn_3P_3$ (Ln= La–Nd, Sm, Gd)," *J. Phys. Soc. Jpn.* **89**, 074707, July 2020, https://doi.org/10.7566/JPSJ.89.074707

[51] L. Wang et al., "Synthesis, structure and magnetism of $RTa_7O_{19}$ (R= Pr, Sm, Eu, Gd, Dy, Ho) with perfect triangular lattice," *J. Alloys Compd.* **937**, 168390, March 2023, https://doi.org/10.1016/j.jallcom.2022.168390

[52] A. N. Rutherford et al., "Magnetic properties of $RE_2O_2CO_3$ (RE = Pr, Nd, Gd, Tb, Dy, Ho, Er, Yb) with a rare earth-bilayer of triangular lattice," *arXiv:2407.08606*, July 2024, https://doi.org/10.48550/arXiv.2407.08606

[53] S. Guo et al., "$NaBaR(BO_3)_2$ (R= Dy, Ho, Er and Tm): structurally perfect triangular lattice materials with two rare earth layers," *Mater. Res. Express* **6**, 106110, October 2019, https://doi.org/10.1088/2053-1591/ab3e2e

[54] S. Zheng et al., "Interplay between crystal field and magnetic anisotropy in the triangular-lattice antiferromagnet $NaTmTe_2$," *Phys. Rev. B* **109**, 075159, February 2024, https://doi.org/10.1103/PhysRevB.109.075159

[55] T. Arh et al., "The Ising triangular-lattice antiferromagnet neodymium heptatantalate as a quantum spin liquid candidate," *Nat. Mater.* **21**, 416, April 2022, https://doi.org/10.1038/s41563-021-01165-y

[56] R. Moessner et al., "Two-Dimensional Periodic Frustrated Ising Models in a Transverse Field," *Phys. Rev. Lett.* **84**, 4457, May 2000, https://doi.org/10.1103/PhysRevLett.84.4457

[57] G. H. Wannier, "Antiferromagnetism. The Triangular Ising Net," *Phys. Rev.* **79**, 357, July 1950, https://doi.org/10.1103/PhysRev.79.357

[58] G. Bastien et al., "A frustrated antipolar phase analogous to classical spin liquids," *Adv.*

*Mater*. **36**, 2410282 (2024), https://doi.org/10.1002/adma.202410282

[59] A. Kahn et al., "Preparation, structure, optical, and magnetic properties of lanthanide aluminate single crystals ($LnMgAl_{11}O_{19}$)," *J. Appl. Phys.* **52**, 6864, November 1981, https://doi.org/10.1063/1.328638

[60] A. Scheie, "PyCrystalField: software for calculation, analysis and fitting of crystal electric field Hamiltonians," *J. Appl. Crystallogr*. **54**, 356, February 2021, https://doi.org/10.1107/S1600576720016130

[61] D. Yamamoto, G. Marmorini, and I. Danshita, "Quantum Phase Diagram of the Triangular-Lattice $XXZ$ Model in a Magnetic Field," Phys. Rev. Lett. 112, 127203, March 2014, https://doi.org/10.1103/PhysRevLett.112.127203

**Supplementary Information for:**

# Induced-quantum magnetism on a triangular lattice of non-Kramers ions in $PrMgAl_{11}O_{19}$

S. Kumar,[1,2], M. Klicpera,[1] A. Eliáš,[1] M. Kratochvílová,[1] A. Kancko,[1] C. Correa,[3] K. Załęski,[4] M. Śliwińska-Bartkowiak,[2] R. H. Colman,[1] and G. Bastien[1*]

[1] *Charles University, Faculty of Mathematics and Physics, Department of Condensed Matter Physics, Prague, Czech Republic*

[2] *Adam Mickiewicz University, Faculty of Physics and Astronomy, Department of Experimental Physics of Condensed Phase, Poznan, Poland*

[3] *Institute of Physics of the Czech Academy of Sciences, Na Slovance, Prague, Czech Republic*

[4] *Adam Mickiewicz University, NanoBioMedical Centre, Poznan, Poland*

**Corresponding author:* sonu.kumar@matfyz.cuni.cz

## Structural analysis

Single crystal X-ray diffraction (SCXRD) was performed at 95 K on a Rigaku SuperNova diffractometer equipped with an Atlas S2 CCD detector, using a mirror-collimated Mo Kα ($\lambda$ = 0.71073 Å) radiation from a micro-focus sealed tube. Diffraction data were integrated using CrysAlis Pro [1] with an empirical absorption correction using spherical harmonics [2] combined with an analytical numeric absorption correction based on Gaussian integration over a multifaceted crystal model implemented in the SCALE3 ABSPACK scaling algorithm. The structure was solved by charge flipping using the program Superflip [3] and refined by full-matrix least-squares on $F^2$ in Jana2020. [4] Structural graphics were created using Jana2020 and Vesta. [5]

The crystal structure of $PrMgAl_{11}O_{19}$ obtained by SCXRD at 95 K is hexagonal, space group *P6₃/mmc* (#194), $Z = 2$, and with unit cell parameters $a$ = 5.5742(3) Å and $c$ = 21.8708(13) Å, in line with the expected magnetoplumbite structure. Due to the identical electronic configuration of $Mg^{2+}$ and $Al^{3+}$ and, therefore, very similar X-ray atomic scattering factors, it is very difficult to distinguish the site of $Mg^{2+}$ ions site via SCXRD. Therefore, an initial model of $PrAl_{12}O_{19}$ was assumed, giving $GOF_{obs}$ = 4.78% and $R_{obs}$ = 5.18%. Allowing the Pr(1) occupancy to refine led to a significant improvement of the model, with *occ*(Pr1) = 0.869(3), $GOF_{obs}$ = 2.32%, and $R_{obs}$ = 2.87%, indicating a 13% $Pr^{3+}$ deficiency. This was initially proposed in $LaMgAl_{11}O_{19}$ [6] and confirmed with neutron diffraction in $CeMgAl_{11}O_{19}$ [7] suggesting that $Mg^{2+}$ is shared with $Al^{3+}$ on the Al(4) site. Splitting the Al(4) site between Mg and Al with fixed 0.5/0.5 occupancies, with harmonic ADPs and coordinates constrained to be equal, the quality of the refinement practically stays unchanged - $GOF_{obs}$ = 2.35% and $R_{obs}$ = 2.86%. Individually placing the Mg ion on each of the remaining Al sites did not lead to an improvement of the model. In the magnetoplumbite structure, the Al(5) ion is often distributed between two off-centered positions in the oxygen bipyramid, with the *z*-coordinate refined to *z*(Al5) = 0.2420(10), resulting in the off-centering distance δ = 2(0.25 - *z*(Al5))*c* = 0.31 Å. As was found previously, an additional improvement of the model was reached by adding a Pr(2) atom to the 6*h* site at a distance of 0.847 Å from Pr(1) located on the 2*d* site [8]. Inspection of the Fourier difference maps in the *ab*-plane cut at *z* / *c* = 0.25 (Figure S1) indicates a partially occupied atomic site (density 1.35 eÅ$^{-3}$ and charge 0.50 e). Inclusion of this positionally disordered (off-centered) Pr ion results in a non-negligible improvement of the agreement factors and yields final $GOF_{obs}$ = 1.54% and $R_{obs}$ = 1.94%, and refined occupancies occ(Pr1) = 0.837(8) and occ(Pr2) = 0.016(3). This solution gives the final formula $Pr_{0.884}MgAl_{11}O_{19}$, suggesting an ~12% total $Pr^{3+}$ deficiency. More details can be found in the deposited CIF files at the Cambridge Crystallographic Data Centre (CCDC), code 2441166 at www.ccdc.cam.ac.uk.

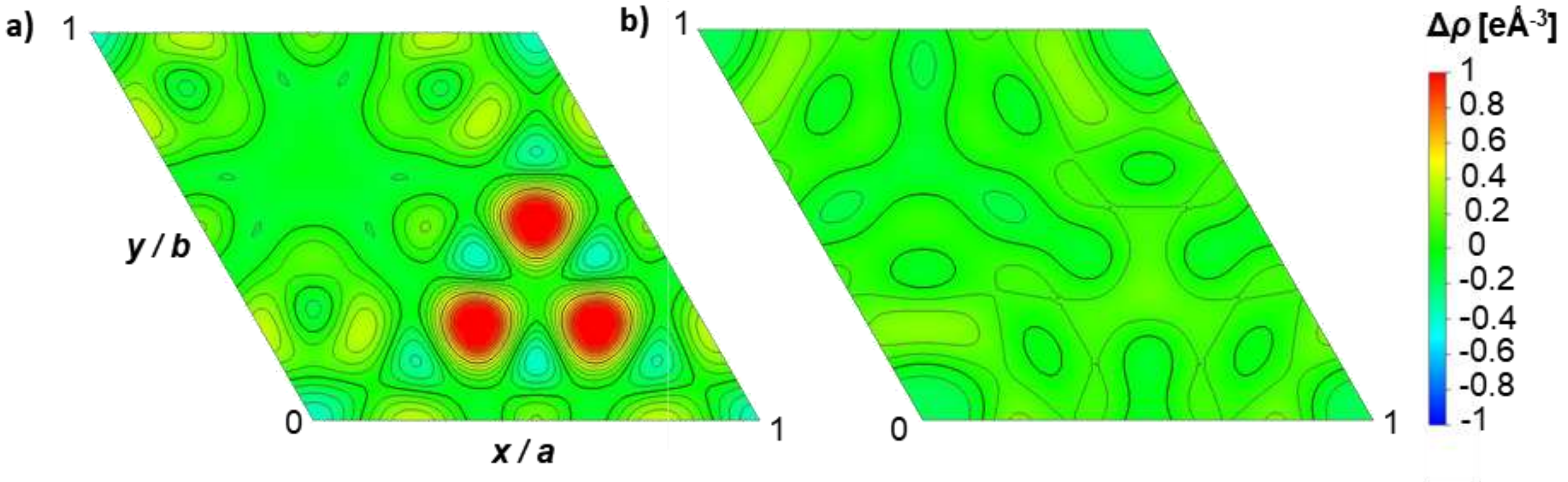

Figure S1. Fourier difference electron density maps within the z / c = ¼ plane, a) before, and b) after the inclusion of the partially occupied, and positionally disordered Pr(2) site. The need for this positional disorder is clearly evidenced by the strong electron pockets seen before the inclusion of Pr(2), as well as the resulting flat difference-electron-density once this positional disorder is properly accounted for within the structural model.

Table S2. Single crystal structure solution results summary.

$Pr_{0.884}MgAl_{11}O_{19}$

| Crystal System & Space Group: | | | | | Crystal size: |
|---|---|---|---|---|---|
| hexagonal, $P6_3/mmc$ (#194, setting 1) | $a$ = 5.5742(3) Å | $c$ = 21.8708(13) Å | $V$ = 588.52(6) Å$^3$ | $Z$ = 2 | 45x31x5 µm |

| Radiation: | Reflections collected / unique / used: | Final indices: | Index ranges: | Maximum difference: | Density: |
|---|---|---|---|---|---|
| X-ray tube: Mo Kα<br>($\lambda$ = 0.71073 Å) | 9053 / 365 / 324<br>$R_{int}$ = 4.1 % | $GOF_{obs}$ = 1.54%<br>$R_{obs}$ = 1.94 %<br>$wR_2$ = 4.47 %<br>Parameters: 46 | $-7 \le h \le 7$<br>$-7 \le k \le 7$<br>$-29 \le l \le 27$<br>$\vartheta$ range:<br>3.73° – 29.68° | Peaks 0.33 eÅ$^{-3}$<br>Holes -0.43 eÅ$^{-3}$ | $\rho$ = 4.2307 g.cm$^{-3}$<br>Absorption coefficient:<br>$\mu$ = 4.699 mm$^{-1}$ |

| $T$ (K) | Atom | Site | $x$ | $y$ | $z$ | Occ. | $U_{ij}$ (×10$^4$ Å$^2$) | | | | | | |
|---|---|---|---|---|---|---|---|---|---|---|---|---|---|
| | | | | | | | $U_{11}$ | $U_{22}$ | $U_{33}$ | $U_{12}$ | $U_{13}$ | $U_{23}$ | $U_{eq}$ |
| 95 K | Pr(1) | 2$d$ | ⅓ | ⅔ | ¾ | 0.837(8) | 69(5) | $U_{11}$ | 28(3) | 0.5$U_{11}$ | 0 | 0 | 55(3) |
| | Pr(2) | 6$h$ | 0.732(4) | 0.464(4) | ¼ | 0.016(3) | 70(60) | 70(60) | 70(60) | 0 | 0 | 0 | 70(60) |
| | Al(1) | 12$k$ | 0.16775(7) | 0.33549(14) | 0.60845(4) | 1 | 34(3) | 38(4) | 42(5) | 0.5$U_{22}$ | 2(2) | 2$U_{13}$ | 38(3) |
| | Al(2) | 4$f$ | ⅓ | ⅔ | 0.19019(6) | 1 | 35(4) | $U_{11}$ | 30(7) | 0.5$U_{11}$ | 0 | 0 | 34(4) |
| | Al(3) | 2$a$ | 0 | 0 | 0 | 1 | 37(6) | $U_{11}$ | 30(12) | 0.5$U_{11}$ | 0 | 0 | 35(5) |
| | Al(4) | 4$f$ | ⅓ | ⅔ | 0.02745(7) | 0.5 | 19(4) | $U_{11}$ | 35(8) | 0.5$U_{11}$ | 0 | 0 | 25(4) |
| | Mg(1) | 4$f$ | ⅓ | ⅔ | 0.02745(7) | 0.5 | 19(4) | $U_{11}$ | 35(8) | 0.5$U_{11}$ | 0 | 0 | 25(4) |
| | Al(5) | 4$e$ | 0 | 0 | 0.24240(10) | 0.5 | 60(7) | $U_{11}$ | 170(10) | 0.5$U_{11}$ | 0 | 0 | 100(3) |
| | O(1) | 6$h$ | 0.1815(2) | 0.3630(4) | ¼ | 1 | 108(10) | $U_{11}$ | 39(14) | 86(11) | 0 | 0 | 71(9) |
| | O(2) | 12$k$ | 0.50502(17) | 0.01004(17) | 0.15101(9) | 1 | 32(6) | $U_{11}$ | 61(11) | 12(7) | 8(4) | -$U_{13}$ | 44(6) |
| | O(3) | 4$e$ | 0 | 0 | 0.15088(15) | 1 | 38(9) | $U_{11}$ | 64(18) | 0.5$U_{11}$ | 0 | 0 | 47(8) |
| | O(4) | 12$k$ | 0.15271(16) | 0.30542(32) | 0.05330(9) | 1 | 49(6) | $U_{11}$ | 58(10) | 14(7) | -9(4) | -$U_{13}$ | 57(6) |
| | O(5) | 4$f$ | ⅓ | ⅔ | 0.55737(16) | 1 | 27(9) | $U_{11}$ | 55(16) | 0.5$U_{11}$ | 0 | 0 | 36(8) |

**Point-charge crystal electric field analysis**

The crystal field Hamiltonian was estimated utilizing the Python-based calculation tool PyCrystalField (version 2.3.10) [9]. As input, the crystallographic information file (cif) for the above described single crystal structure solution was used. The magnetic atom was defined as $Pr^{3+}$, and surrounding ligand oxygens were confirmed as $O^{2-}$. The calculation was extended only to the first nearest neighbors (i.e. the $PrO_{12}$ polyhedral unit). The [1,-1,0], [1,1,0], and [0,0,1] crystallographic directions were used to define the local $x$, $y$, and $z$, directions for the calculation of the crystal field Hamiltonian, respectively. The calculated point-charge model can be represented by the Hamiltonian $\mathcal{H}_{\text{CEF}} = \sum_{n,m} B_n^m O_n^m$, with only $B_2^0$, $B_4^0$, $B_6^0$, and $B_6^6$ as non-zero Steven's operators. The determined coefficients are -0.85659083, -0.00418386, -0.00031973, and -0.00239255, respectively. The resulting eigenvalues and eigenvectors are shown in Table S3.

Table S3. Calculated Eigenvectors and Eigenvalues for $Pr^{3+}$ within $PrMgAl_{11}O_{19}$

| Eigenvalues (meV) | Eigenvectors $\|-4\rangle$ | $\|-3\rangle$ | $\|-2\rangle$ | $\|-1\rangle$ | $\|0\rangle$ | $\|1\rangle$ | $\|2\rangle$ | $\|3\rangle$ | $\|4\rangle$ |
|---|---|---|---|---|---|---|---|---|---|
| 0.00000 | -0.989 | 0.0 | 0.0 | 0.0 | 0.0 | 0.0 | -0.148 | 0.0 | 0.0 |
| 0.00000 | 0.0 | 0.0 | -0.148 | 0.0 | 0.0 | 0.0 | -0.989 | 0.0 | 0.0 |
| 29.88539 | 0.0 | 0.707 | 0.0 | 0.0 | 0.0 | 0.0 | 0.0 | 0.707 | 0.0 |
| 31.22192 | -0.148 | 0.0 | 0.0 | 0.0 | 0.0 | 0.0 | 0.989 | 0.0 | 0.0 |
| 31.22192 | 0.0 | 0.0 | -0.989 | 0.0 | 0.0 | 0.0 | 0.0 | 0.0 | 0.148 |
| 41.69044 | 0.0 | 0.0 | 0.0 | -1.0 | 0.0 | 0.0 | 0.0 | 0.0 | 0.0 |
| 41.69044 | 0.0 | 0.0 | 0.0 | 0.0 | 0.0 | -1.0 | 0.0 | 0.0 | 0.0 |
| 41.94385 | 0.0 | -0.707 | 0.0 | 0.0 | 0.0 | 0.0 | 0.0 | -0.707 | 0.0 |
| 50.46090 | 0.0 | 0.0 | 0.0 | 0.0 | -1 | 0.0 | 0.0 | 0.0 | 0.0 |

The resulting ground-state $g$-tensor remains Ising-like: $\begin{bmatrix} 0.0 & 0.0 & 0.0 \\ 0.0 & 0.0 & 0.0 \\ 0.0 & 0.0 & 6.19 \end{bmatrix}$.

**Schottky Analysis**

The magnetic specific heat $Cm/T$ vs $T$ of $PrMgAl_{11}O_{19}$ exhibits a double-hump structure (Fig. S2.a), indicating two distinct energy splittings. This behavior is attributed to two inequivalent $Pr^{3+}$ sites in the crystal structure, each subject to a unique local CEF. To analyze these features, the data were modeled using a two-gap Schottky anomaly of the form:

$$\frac{C_m}{T}(T) = \frac{R}{T}\sum_{j=1}^{2} f_j \left(\frac{\Delta_j}{k_B T}\right)^2 \frac{Exp\left(\frac{\Delta_j}{k_B T}\right)}{\left(1 + Exp\left(\frac{\Delta_j}{k_B T}\right)\right)^2}$$

where $f_1$ and $f_2$ represent the fractions of ions contributing to each Schottky term, and $\Delta_1$ and $\Delta_2$ are the corresponding energy gaps. At zero field, the smaller gap $\Delta_1$=1.26meV corresponds to the lower-energy Schottky peak near 5 K, while the larger gap $\Delta_2$=5.16meV produces the higher-energy shoulder near 13 K. The fractional occupancies $f_1 \approx 0.42$ and $f_2 \approx 0.58$ align with the two crystallographically distinct $Pr^{3+}$ sites identified in XRD refinements.

Under an applied magnetic field, $\Delta_1$ exhibits a field dependence (Fig. S2.b), where the zero-field gap is added quadratically to the Zeeman energy [10, 11] as:

$$\Delta(H) = \sqrt{{\Delta_0}^2 + (\mu B \mu_0 H)^2}\ , \qquad \text{(S.1)}$$

yielding an effective $g_c$=5.6, consistent with ESR (= 5) [7] and CW analysis (5.1). Below 7 T, the two hump-like features become more discernible as the field reduces the overlap between the Schottky anomalies, though they remain partially merged. At high fields ($\mu_0 H \geq 7$T), the anomalies converge into a single broad peak due to competing Zeeman and exchange effects. In contrast, $\Delta_2$ shows minimal field sensitivity, with an extracted $g_c \approx 9.8$. This anomalously high value may arise from exchange interactions and local CEF distortions, which enhance the effective field beyond the single-ion Zeeman limit [12], or it may also arise due to bad resolution of fit for $\Delta_2$.

At low field (0-3 T), deviations from the two-gap fit are more pronounced, reflecting perturbations from magnetic correlations and disorder-induced broadening of the Schottky anomalies, also noted by field dependence fit of Δ (Fig. S2 (b)). These deviations suggest additional interactions in the CEF environment. The extracted *g*-factors = 9.63, align with prior studies on frustrated magnets, such as $TmMgGaO_4$ where exchange-enhanced fields produce renormalized *g*-factors exceeding theoretical maxima [12]

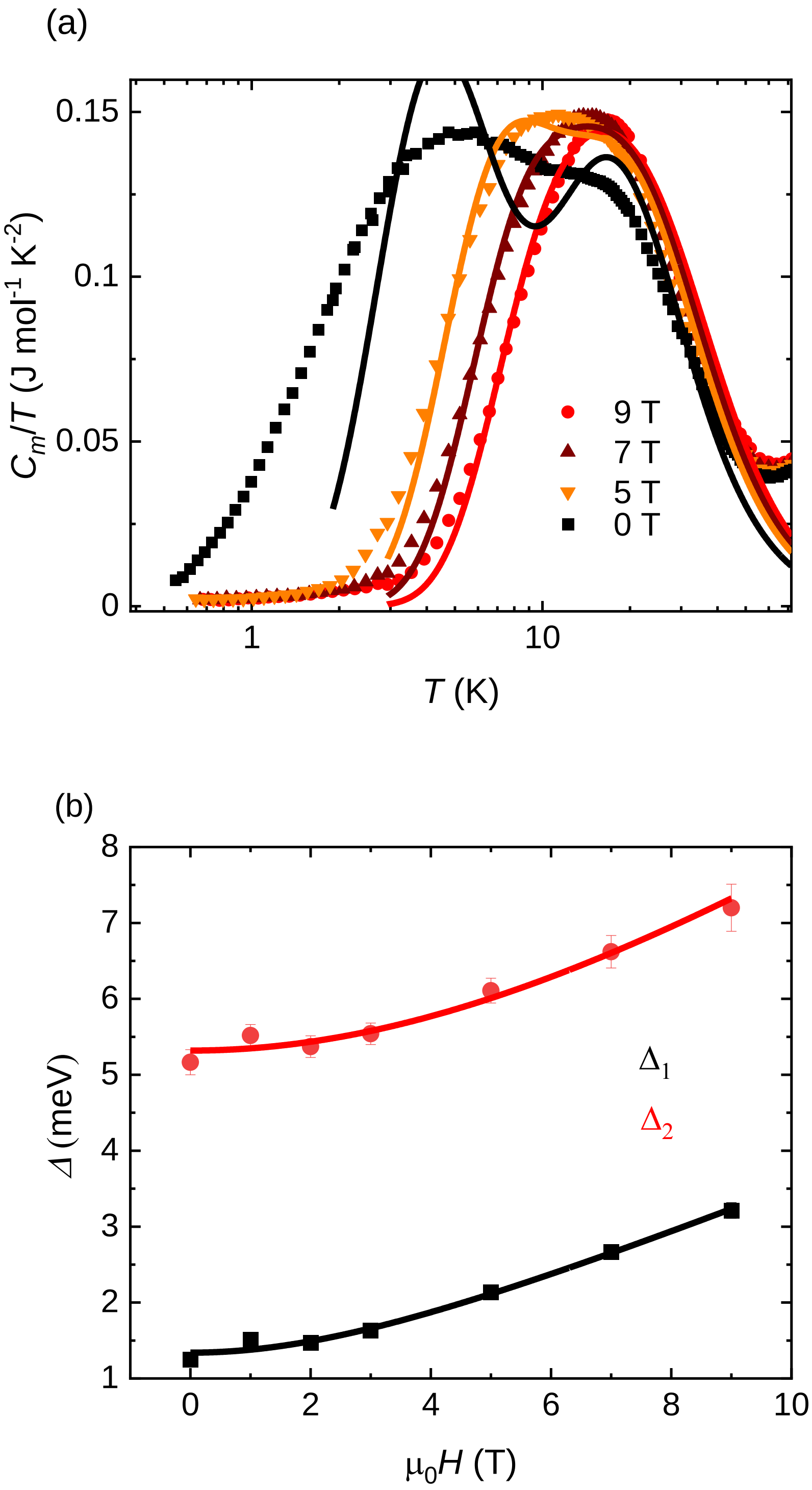

Fig. S2: (a) $Cm/T$ as a function of temperature for $PrMgAl_{11}O_{19}$ at selected magnetic fields. The solid lines represent the best fits to the two-gap Schottky model, accounting for two inequivalent $Pr^{3+}$ sites with distinct crystal electric field (CEF) splittings. (b) Field dependence of the energy gaps $\Delta_1$ and $\Delta_2$ extracted from the Schottky fits. Solid lines represent the field dependence fit of $\Delta$ using equation (S.1).

**References**

[1] CrysAlisPro software (version 1.171.43.143a, Rigaku Oxford Diffraction, 2024).

[2] R. C. Clark and J. S. Reid, “The analytical calculation of absorption in multifaceted crystals,” *Acta Crystallogr. Sect. A*, vol. 51, no. 6, pp. 887–897, 1995, doi: 10.1107/S0108767395007367.

[3] L. Palatinus and G. Chapuis, “SUPERFLIP - A computer program for the solution of crystal structures by charge flipping in arbitrary dimensions,” *J. Appl. Crystallogr.*, vol. 40, no. 4, pp. 786–790, 2007, doi: 10.1107/S0021889807029238.

[4] V. Petříček, L. Palatinus, J. Plášil, and M. Dušek, “Jana2020 – a new version of the crystallographic computing system Jana,” *Zeitschrift für Krist. - Cryst. Mater.*, vol. 238, no. 7–8, pp. 271–282, Jul. 2023, doi: 10.1515/zkri-2023-0005.

[5] K. Momma and F. Izumi, “VESTA: a three-dimensional visualization system for electronic and structural analysis,” *J. Appl. Crystallogr.*, vol. 41, no. 3, pp. 653–658, Jun. 2008, doi: 10.1107/S0021889808012016.

[6] M. Gasperin et al., “Influence of $M^{2+}$ ions substitution on the structure of lanthanum hexaaluminates with magnetoplumbite structure,” *J. Solid State Chem.*, vol. 54, no. 1, pp. 61–69, Aug. 1984, doi: 10.1016/0022-4596(84)90188-5.

[7] B. Gao et al., “Spin Excitation Continuum in the Exactly Solvable Triangular-Lattice Spin Liquid $CeMgAl_{11}O_{19}$,” arXiv:2408.15957, 2024.

[8] Y. Cao et al., “Synthesis, disorder and Ising anisotropy in a new spin liquid candidate $PrMgAl_{11}O_{19}$,” *Mater. Futur.*, vol. 3, no. 3, 2024, doi: 10.1088/2752-5724/ad4a93.

[9] A. Scheie, “PyCrystalField: Software for calculation, analysis, and fitting of crystal electric field Hamiltonians,” *J. Appl. Cryst.*, vol. 54, no. 1, pp. 3–10, Feb. 2021, doi: 10.1107/S160057672001554X.

[10] P. Thalmeier and A. Akbari, “Induced quantum magnetism in crystalline electric field singlet ground state models: Thermodynamics and excitations,” *Phys. Rev. B*, vol. 109, no. 11, p. 115110, Mar. 2024, doi: 10.1103/PhysRevB.109.115110.

[11] J. Nagl et al., “Excitation spectrum and spin Hamiltonian of the frustrated quantum Ising magnet $Pr_3BWO_9$,” *Phys. Rev. Res.*, vol. 6, no. 2, p. 023267, Apr. 2024, doi: 10.1103/PhysRevResearch.6.023267.

[12] Shen, Y., Liu, J., Li, Y., et al. (2019). Intertwined dipolar and multipolar order in the

triangular-lattice magnet $TmMgGaO_4$. *Nature Communications*, 10(1), 4530.
https://doi.org/10.1038/s41467-019-12410-3